\documentclass[12pt]{article}
\usepackage{hyperref}
\usepackage{amsmath}
\usepackage{graphicx}
\usepackage{amssymb} 
\usepackage{xcolor}
\definecolor{myblue}{rgb}{0.14,0.11,0.49}
\definecolor{myred}{rgb}{0.74,0.22,0.15}

\definecolor{mygreen}{rgb}{0.05,0.52,0.42}
\definecolor{myyellow}{rgb}{0.96,0.92,0.13}
\definecolor{myorange}{rgb}{1,0.61,0.36}
\definecolor{mypurple}{rgb}{0.71,0.02,1}

\def\be{\begin{equation}}
\def\ee{\end{equation}}
\def\bea{\begin{eqnarray}}
\def\eea{\end{eqnarray}}
\def\dd{\mathrm{d}}

\date{}

\title{{\bf A non-uniqueness problem of the Dirac theory in a curved spacetime}}

\author{
Mayeul Arminjon\,$^{1}$ and Frank Reifler\,$^2$\\
$^1$ \small\it Laboratory ``Soils, Solids, Structures, Risks'', 3SR\\ \small\it (CNRS and Universit\'es de Grenoble: UJF, Grenoble-INP),\\
\small\it BP 53, F-38041 Grenoble cedex 9, France.\\
$^2$ \small\it Lockheed Martin Corporation, MS2 137-205,\\ 
\small\it 199 Borton Landing Road, Moorestown, New Jersey 08057, USA.
} 

\begin{document}
\maketitle
\begin{abstract}
\noindent The Dirac equation in a curved spacetime depends on a field of coefficients (essentially the Dirac matrices), for which a continuum of different choices are possible. We study the conditions under which a change of the coefficient fields leads to an equivalent Hamiltonian operator H, or to an equivalent energy operator E. We do that for the standard version of the gravitational Dirac equation, and for two alternative equations based on the tensor representation of the Dirac fields. The latter equations may be defined when the spacetime is four-dimensional, noncompact, and admits a spinor structure. We find that, for each among the three versions of the equation, the vast majority of the possible coefficient changes do not lead to an equivalent operator H, nor to an equivalent operator E, whence a lack of uniqueness. In particular, we prove that the Dirac energy spectrum is not unique. This non-uniqueness of the energy spectrum comes from an effect of the choice of coefficients, and applies in any given coordinates.

\end{abstract}

\section{Introduction}\label{Introduction}
Dirac's equation for a relativistic spin-half particle was originally proposed for a Minkowski spacetime. Dirac's original equation is limited to Cartesian coordinates, and the coefficients which enter that equation---the Dirac matrices---are constant. In a curved spacetime, or for non-Cartesian coordinates in a Minkowski spacetime, Dirac's original equation does not apply any more, and is usually replaced \cite{BrillWheeler1957+Corr,deOliveiraTiomno1962,ChapmanLeiter1976} by the equation proposed independently by Weyl \cite{Weyl1929b} and by Fock \cite{Fock1929b}, hereafter the ``Dirac-Fock-Weyl" equation ({\it DFW} for short). This equation has been investigated in situations of physical interest, {\it e.g.} in uniformly rotating coordinates in a Minkowski spacetime ({\it e.g.} \cite{ChapmanLeiter1976}), in uniformly accelerating coordinates in a Minkowski spacetime or in a static, or stationary, weak gravitational field ({\it e.g.} \cite{deOliveiraTiomno1962,SoffelMullerGreiner1977,HehlNi1990,VarjuRyder1998,Obukhov2001,A38,Boulanger-Spindel2006}), in a Robertson-Walker spacetime \cite{Villalba1990
}, etc. While the currently observable effects of the gravitational field on quantum particles (such as neutrons or atoms) can still be described by using the non-relativistic Schr\"odinger equation in the Newtonian gravity potential, one may expect that, in a not-too-distant future, improvements in the experimental accuracy should allow one to check the combined effects of relativistic gravity and relativistic quantum mechanics. This will be a check of the way in which we conceive the coupling between gravity and the quantum, including a check of the way we write the quantum wave equations in a curved spacetime.\\

In the DFW equation, ``curved" Dirac matrices $\gamma ^\mu(X) $ most generally vary from one spacetime point $X$ to another one, even though they are expressed linearly as functions of a fixed set of constant Dirac matrices, say $\gamma ^{\natural \alpha }$, through a matrix of coefficients $a^\mu_{\ \,\alpha}(X)$ made with the components, in the coordinate basis $(\partial _\mu )$, of an orthonormal tetrad field $u_\alpha =a^\mu_{\ \,\alpha}\,\partial_\mu$: \footnote{\
Here and in the whole of the paper, we are using ``early" Greek letters $\alpha,\beta,...$ for tetrad (or frame) indices corresponding to the tangent bundle TV, whereas ``middle" Greek letters $\mu,\nu,...$  are denoting coordinate indices.
}
\be \label{flat-deformed}
  \gamma ^\mu(X) = a^\mu_{\ \,\alpha}(X)  \ \gamma ^{ \natural \alpha}.
\ee 
In a given Lorentzian spacetime $(\mathrm{V},g_{\mu \nu })$, there is a continuum of possible choices for the tetrad field and its matrix $a(X)\equiv (a^\mu _{\ \,\alpha})$---this has been emphasized recently by other authors also \cite{Gorbatenko-Gorbatenko2008,McLenaghan-et-al2008}. The tetrad field is assigned to satisfy the orthonormality condition:
\be\label{compatible-anticom}
a^\mu _{\ \,\alpha} \,a^\nu _{\ \,\beta } \,\eta ^{\alpha \beta }=g^{\mu \nu }, \qquad (g^{\mu \nu})\equiv(g_{\mu \nu})^{-1}.
\ee
[Here $(\eta ^{\alpha \beta  })\equiv (\eta _{\alpha \beta })^{-1}=(\eta _{\alpha \beta })\equiv \mathrm{diag}(1,-1,-1,-1)$.] Moreover, any set $(\gamma ^{\natural \alpha })$ of ``flat" Dirac matrices [{\it i.e.,} any constant solution of the anticommutation relation (\ref{Clifford}) below with $g^{\mu \nu }=\eta ^{\mu \nu }$] enjoys exactly the same status. Hence, none should be preferred over another one \cite{A40}, although in the literature the choice of the set $(\gamma ^{\natural \alpha })$ is restricted, unnecessarily as we will see in Section \ref{Hamiltonian-modified}, by a hermiticity condition \cite{BetheJackiw1968,Ryder1996}.\\

As stated above, in Eq. (\ref{flat-deformed}) there is a continuum of possible choices for the field $\gamma ^\mu $. This is also true for the two alternative extensions of the Dirac equation to a curved spacetime, which have been proposed recently \cite{A39}.
Each of these two alternative Dirac equations in a curved spacetime is based on the {\it tensor representation of the Dirac} ({\it TRD}) field \cite{A39,A37}, in which the wave function is a four-vector and the set of the components of the four $\gamma ^\mu $ matrices builds a third-order tensor.
\footnote{\
In a Minkowski spacetime in Cartesian (or affine) coordinates, the TRD theory with constant Dirac matrices has been proved \cite{A40} to be quantum-mechanically fully equivalent to the genuine Dirac theory. See also Subsect. 1.1 in Ref. \cite{A39} for a summary of the argument. Like the solutions of the genuine Dirac equation, the single particle TRD solutions have only two spin polarizations (up and down) which makes them spin-half wave functions. Moreover, just as with the genuine Dirac equation, it is straightforward to extend the single particle theory to a canonical second quantized fermion theory (see Subsect. 4.3 of Ref. \cite{Ryder1996}), with the fermion field operator built on the normalized single particle and antiparticle TRD solutions.  
}\\

In a Minkowski spacetime, the Dirac wave function transforms under the spin group as the coordinates are Lorentz-transformed. However, it is well known since the work of Fock and Weyl \cite{Weyl1929b,Fock1929b} that, on changing the coordinates, transforming the Dirac wave function under the spin group is not an option in a curved spacetime or even in a flat spacetime with affine coordinates. Thus, there are only two possibilities in a curved spacetime. Either the Dirac wave function transforms as a quadruplet of four scalar fields under coordinate transformations as in DFW, or it transforms as a complex four-vector field as in TRD. As discussed in Sect. \ref{framework}, the latter four-vector transformation is possible whenever the spacetime manifold is four-dimensional, noncompact, and admits a spinor structure. In a flat spacetime with affine coordinates, and constant Dirac matrices, TRD and DFW have been shown to be equivalent \cite{A40}. From the Lagrangian in Subsect. \ref{Lagrangian}, we can show in a straightforward manner that this equivalence gives rise to equivalent fermion quantum field theories in a flat spacetime with affine coordinates.\\  

However, to our knowledge, there has been hardly any study of the possible effect of the choice of the field $\gamma ^\mu $, even for the standard, DFW theory. Moreover, when one uses a general set of Dirac matrices, one needs a matrix $A$, called the {\it hermitizing matrix,} that also becomes a field of matrices $A(X)$ in a curved spacetime. So the Dirac equation in a given spacetime depends on the fields $(\gamma ^\mu ,A)$, which we call the {\it coefficient fields} of the Dirac equation. In a recent work \cite{A42}, the quantum mechanics associated with these three gravitational Dirac equations---the standard, DFW equation, and the two TRD equations---has been studied with regard to conservation of the probability current, Hilbert space scalar product, and hermiticity of the Hamiltonian operator. In particular, the hermiticity condition has been derived in a general spacetime and it has been shown that, {\it for DFW, the hermiticity of the Dirac Hamiltonian is not preserved under all admissible changes of the coefficient fields $(\gamma ^\mu ,A)$.}\\

This surprising result indicates that the standard version of the gravitational Dirac equation has a non-uniqueness problem. The aim of the present work is, therefore, to begin the systematic study of the non-uniqueness problem of the Dirac theory in a curved spacetime. The result just recalled points to a non-uniqueness of the Hamiltonian theory, thus, possibly, of the energy spectrum associated with the Dirac equation in a curved spacetime. In a time-dependent metric, the Hamiltonian operator, that is the generator of the time evolution, is not necessarily Hermitian and does not generally coincide with the energy operator, associated with the field energy. It is the energy operator, not the Hamiltonian, which is derived canonically from the Lagrangian: see Ref. \cite{Leclerc2006}, and Sect. \ref{Equivalent-E-Operators} below.  Hence, in the present work, we focus on the non-uniqueness problem of the Hamiltonian and energy operators, both having physical relevance. We ask successively: When does a change of the coefficient fields lead to an equivalent Hamiltonian operator? Or to an equivalent energy operator? In any given coordinate system is the energy spectrum unique?  We shall investigate these questions for the DFW theory as well as for the alternative, TRD, theory.  In particular, in Sect. \ref{Non-uniquenessProof} we prove that, in any given coordinate system, the spectrum of the energy operator for the free Dirac equation is not unique in both DFW and TRD, but depends on the tetrad field or set of coefficient fields that is chosen.\\

While it is expected that the energy spectrum would depend on the choice of the time coordinate, it is unexpected to discover that, in any given coordinate system (in particular, with any given time coordinate), the energy spectrum varies with the choice of the coefficient fields $(\gamma ^\mu ,A)$. Currently, there is no preferred way to choose these coefficient fields in a general curved spacetime.  The non-uniqueness of the energy spectrum is significant because it is a fundamental axiom in quantum mechanics that the spectrum of the energy operator at a given time $t$ restricts the allowed values of energy that will be observed in experiments at time $t$. This axiom, first enunciated by Max Born, is postulated to be true of both time-dependent and time-independent quantum theories. Consequences of Born's axiom together with the non-uniqueness of the energy spectrum in any given coordinate system are discussed briefly at the end of Sect. \ref{Conclusion}. We show that the energy shifts due to different ``reasonable'' choices of the coefficient fields $(\gamma^\mu,A)$ are very small, similar in magnitude to the Mashhoon energy term for a rotating coordinate system in a Minkowski spacetime.  Such small energy shifts, described in Sect. \ref{Conclusion}, are too small to observe currently, but may be detected in the future with improved experimental techniques \cite{Mashhoon1988}

\section{Framework}\label{framework}

\paragraph{i) First quantized Dirac theory.} There is a long list of physical effects whose interpretation uses the first quantized Dirac theory and does not require second quantization---e.g., the Stern-Gerlach, COW, Sagnac, and Mashhoon effects. Moreover, it is well known that the problems of quantum field theory in a curved spacetime do not all reside within quantum field theory itself, since some of the most challenging problems for quantum fields in a curved spacetime involve difficulties in defining the single particle states, along with their Hilbert space structure, and the operators that act on them.  These difficulties with single particle states, which are used to define the field operators, the vacuum, and the occupation states, are not resolved by the mere formalism of quantum field theory. Among these are the difficulties associated with the energy spectrum of single particle states in a curved spacetime.  This is critical, not only for its impact on proposed experiments, but also theoretically for the identification of positive and negative energy states. See for example Refs. \cite{Birrell-Davies1982,Fulling1989}. Therefore, just like Refs. \cite{BrillWheeler1957+Corr}--\cite{Villalba1990}, the present paper addresses the first quantized Dirac theory, which gives the normalized single particle and antiparticle states that define a second quantized fermion field operator in a curved spacetime \cite{Birrell-Davies1982,Fulling1989}.

\paragraph{ii) Trivial topology.}\label{Framework ii} As is well known, a given Lorentzian spacetime $(\mathrm{V},g_{\mu\nu})$ need not admit a spinor structure. However, the situation is greatly simplified if the spacetime is four-dimensional and noncompact. It was proved by Geroch that such a spacetime admits a spinor structure if and only if it admits a global tetrad field \cite{Geroch1968}. In that case, both spinor bundles and tangent bundles are trivializable \cite{Geroch1968,Isham1978}. Penrose and Rindler argue that these are the only spacetimes of interest 
\cite{PenroseRindler1986}. In fact, we will assume that the spacetime manifold V is diffeomorphic to ${\sf R}^4$. The non-uniqueness results proved below do not really depend on this assumption, however it simplifies the discussion. For instance, it allows us to assume that a single coordinate system or chart $\chi:\,X \mapsto (x^\mu)$ is defined over the whole of V.\\

Therefore, some geometrical aspects of the Dirac theory, which are discussed in Ref. \cite{A45}, are not needed in the present study. We note also that the DFW equation is covariant under any admissible local similarity transformation (see Subsect. \ref{Lagrangian}), and hence, except for some complex topologies of the spacetime  \cite{Isham1978}, under any change of the tetrad field in Eq. (\ref{flat-deformed}). However, this fact---that the DFW equation is unique in a topologically simple spacetime---does not prevent the corresponding Hamiltonian and energy operators from being non-unique (in a given reference frame or even in a given coordinate system), as we prove in the present work.\\

\paragraph{iii) Dirac equations in a curved spacetime.} \label{Framework iii} In the present work, we fix the chart $\chi $ (the latter being assigned to cover the spacetime V), and we study the dependence of the Hamiltonian and energy operators on the coefficient fields $(\gamma ^\mu, A)$. We give little detail about the covariance of the definitions introduced under a change of chart, as this covariance is clear, in each case (DFW or TRD), from previous work \cite{A40,A42}. The normal Dirac equations in a curved spacetime $(\mathrm{V},g_{\mu \nu })$, either the standard, DFW equation, or any of the two alternative TRD equations, have the same form: 
\be\label{Dirac-general}
\gamma ^\mu D_\mu\psi=-im\psi,\\
\ee
\noindent where $\gamma ^\mu =\gamma ^\mu (X)$ ($\mu =0,...,3$) is a field of $4\times 4$ complex matrices defined on the spacetime manifold V, satisfying the anticommmutation relation
\be \label{Clifford}
\gamma ^\mu \gamma ^\nu + \gamma ^\nu \gamma ^\mu = 2g^{\mu \nu}\,{\bf 1}_4, \quad \mu ,\nu \in \{0,...,3\} \quad ({\bf 1}_4\equiv \mathrm{diag}(1,1,1,1));
\ee
and where $m$ is the rest-mass of the particle (setting $\hbar=1=c$). In a curved spacetime, $\psi$ is either a quadruplet of four scalar fields (for DFW) or a four-vector field (for TRD).  Thus, for DFW, the field $\psi $ remains invariant after a coordinate change \cite{BrillWheeler1957+Corr,deOliveiraTiomno1962,ChapmanLeiter1976} and, accordingly, the $\gamma ^\mu$ 's transform then as a four-vector. Instead, for TRD, the field $\psi $ is a true four-vector, and the array of components $\gamma^{\mu \rho}_\nu \equiv \left(\gamma^\mu \right)^\rho_{\ \nu}$ builds a $(^2 _1)$ tensor \cite{A39,A40}. If there is a global tetrad field on the spacetime V (as required by Geroch's theorem recalled above), then it is easy to show that one may define a global field of Dirac matrices $\gamma^\mu$  satisfying Eq. (\ref{Clifford}) above, for DFW and for TRD as well (see the appendix in Ref. \cite{A45}). But since we assume that the spacetime is diffeomorphic to ${\sf R}^4$, it is easy to show that indeed it admits a global tetrad field. The three equations differ also in the connection that defines the covariant derivative $D_\mu $. For each of the two TRD equations, this is a connection on the complexification $\mathrm{T}_{\sf C}\mathrm{V}$ of the tangent bundle---namely, the Levi-Civita connection extended to $\mathrm{T}_{\sf C}\mathrm{V}$ for ``TRD--1", and a special affine connection for ``TRD--2" \cite{A39,A42}. For DFW this is the ``spin connection" acting on the trivial bundle $\mathrm{V} \times {\sf C^4}$ \cite{A45}. It is built from the ``spin matrices" $\Gamma _\mu $ \cite{BrillWheeler1957+Corr,deOliveiraTiomno1962,ChapmanLeiter1976},
\be\label{D_mu psi}
D_\mu \psi  \equiv \partial _\mu \psi + \Gamma _\mu \psi.
\ee
The formula above may be used also for TRD--1 and for TRD--2, by defining appropriate matrices $\Gamma _\mu$ from the relevant connection \cite{A42}. The spin matrices $\Gamma _\mu$, of DFW, depend on the set of fields $(\gamma ^\mu) \ (\mu =0,...,3)$. The $\Gamma _\mu$ matrices of TRD--1, and those of TRD--2, do not depend on the fields $(\gamma ^\mu)$.

\paragraph{iv) General hermitizing matrix.} The hermitizing matrix is a nonzero $4 \times 4$ complex matrix $A$ such that \cite{Pauli1933,Pauli1936}:
\be\label{hermitizing-A}
A^\dagger = A, \qquad (A\gamma ^\mu )^\dagger = A\gamma ^\mu \quad \mu =0, ...,3,
\ee
where $M^\dagger\equiv M^{*\,T}$ denotes the Hermitian conjugate of a matrix $M$. Previous work \cite{A40,A42} has proved the existence and uniqueness, up to a real factor $\lambda(X) \ne 0$, of $A=A(X)$, in any spacetime. It has also proved that, if $g_{00}>0$ and the $3\times 3$ matrix $(g_{jk})\ (j,k=1,2,3)$ is negative definite, then the sign of the matrix $A$ can be chosen such that $B\equiv A\gamma ^0$ is a positive definite
\footnote{\ 
Thus, the hermitizing matrix $A(X)$ is unique up to a {\it positive} factor $\lambda(X) > 0$.
}
matrix, which is hermitizing for the alpha matrices:
\be\label{hermitizing-B}
B^\dagger = B, \qquad (B\alpha  ^\mu )^\dagger = B\alpha  ^\mu \quad \mu =0, ...,3,
\ee
where
\be \label{alpha}
\alpha ^0 \equiv \gamma ^0/g^{00}, \qquad \alpha ^j \equiv \gamma ^0\gamma ^j/g^{00}.
\ee
For DFW, the gamma field $\gamma ^\mu (X)$ is given by Eq. (\ref{flat-deformed}). It follows easily from this \cite{A42} that, for DFW, the hermitizing matrix field can be chosen to be:
\be\label{A= A^natural}
\forall X \in \mathrm{V}, \quad A(X) =A^\natural, 
\ee
with $A^\natural \ $ a constant hermitizing matrix for the ``flat" constant Dirac matrices $\gamma ^{ \natural \alpha}$ in Eq. (\ref{flat-deformed}). Moreover, the standard choices, {\it i.e.,} the Dirac, chiral, and Majorana representations of the constant Dirac matrices $\gamma ^{ \natural \alpha}$ in Eq. (\ref{flat-deformed}), are such that $\gamma ^{\natural  0 }$ turns out to be a hermitizing matrix for the set $ (\gamma ^{ \natural \alpha})$. Thus, {\it in most of the literature, the hermitizing matrix field $A$ is just the constant matrix} $\gamma ^{\natural  0 }$. Hence, the reader may assume that $A\equiv \gamma ^{\natural  0 }$ each time that the DFW equation is discussed below, although as mentioned in Refs. \cite{BrillWheeler1957+Corr} and \cite{Audretsch1974}, this is not necessary for DFW. For TRD, we must consider a general hermitizing matrix \cite{A40}, which in a curved spacetime becomes a field \cite{A42}.


\section{Hamiltonians of the Dirac equation and the modified Dirac equation}

\subsection{Hamiltonian of the normal Dirac equation}\label{H-normal}

The Dirac Hamiltonian $\mathrm{H}$, which puts the normal Dirac equation (\ref{Dirac-general}) into Schr\"odinger form:
\be \label{Schrodinger-general}
i \frac{\partial \psi }{\partial t}= \mathrm{H}\psi,\qquad (t\equiv x^0),
\ee
is \cite{A38,A42}:
\be \label{Hamilton-Dirac-normal}
 \mathrm{H} \equiv  m\alpha  ^0 -i\alpha ^j D _j -i\Gamma _0.
\ee
[Here $\Gamma_0=D_0-\partial _0$ is the connection matrix of Eq. (\ref{D_mu psi}).] It is important to note \cite{A42} that, in a given spacetime $(\mathrm{V},g_{\mu \nu })$ and with given coefficient fields $(\gamma ^\mu,A )$, this operator still depends on the coordinate system, or, more exactly, on the {\it reference frame}---the latter being understood here as an equivalence class F of local coordinate systems (charts) on the spacetime V, modulo the purely spatial transformations
\be\label{purely-spatial-change}
x'^0=x^0,\ x'^j=f^j((x^k)).
\ee
As shown in Ref. \cite{A44}, the data of a reference frame F determines a three-dimensional ``space" manifold M, which is the set of the world lines of the observers bound to F---{\it i.e.,} a congruence of time-like curves with spatial coordinates $x^j$  parameterizing the space manifold M and a time coordinate $x^0$ parameterizing the curves. In the present work, the reference frame F is fixed since the chart is fixed as explained \hyperref[Framework iii]{above}.

\subsection{Current conservation: the modified Dirac equation}\label{modified Dirac equation}

The probability current is defined as:
\be \label{J-mu-standard-matrix}
J^\mu =  \psi ^\dagger A\gamma ^\mu \psi.
\ee
This definition is generally-covariant, the current being indeed a four-vector, for TRD and for DFW as well. The current 
(\ref{J-mu-standard-matrix}) is {\it invariant} by a change of the fields $\gamma ^\mu$ through a {\it local similarity transformation} $S=S(X) \in {\sf GL}(4,{\sf C})$. This is a transformation that changes one set $(\gamma ^\mu)$ to another one $(\widetilde {\gamma }^\mu)$, also satisfying  
(\ref{Clifford}), by setting:
\footnote{\
For both the DFW and TRD theories in a non-simply connected spacetime, there may be distinct classes of smooth Dirac matrices $\gamma^\mu$ which do not transform into each other by smooth local similarity transformations \cite{Isham1978}.  However, in this paper, the only changes of the Dirac matrices $\gamma^\mu$ that we consider are induced from smooth local similarity transformations.  This restricts the changes of the $\gamma^\mu$ matrices to be within a single class \cite{Isham1978}.
}
\be \label{similarity-gamma}
\widetilde{\gamma} ^\mu =  S^{-1}\gamma ^\mu S, \quad \mu =0,...,3.
\ee
[Of course, the anticommutation relation (\ref{Clifford}) has to be satisfied by the new gamma matrices $\widetilde{\gamma} ^\mu$ with the {\it same} given metric $g_{\mu \nu }$ as with the starting ones $\gamma ^\mu$. It is easy to check this from Eq. (\ref{similarity-gamma}).]
A hermitizing matrix field for the new fields $\widetilde {\gamma} ^{\mu}$ may be defined thus \cite{A40}:
\be \label{similarity-A}
\widetilde {A} \equiv S^\dagger A S.
\ee
With the corresponding change of the wave function:
\be\label{psitilde=S^-1 psi} 
\widetilde{\psi}=S^{-1}\psi,
\ee
this leads to the invariant relation \cite{A40,A42}
\be \label{J-tilde}
\widetilde {J}^\mu \equiv \widetilde {\psi} ^\dagger \,\widetilde {A}\ \widetilde {\gamma} ^{\mu}\,\widetilde {\psi}= J^\mu.
\ee
Note that the original Dirac theory \cite{BetheJackiw1968,Ryder1996} employed Lorentz coordinate transformations and unitary similarity transformations $S$.  The DFW theory employs general coordinate transformations and local spin transformations. That is to say, local similarity transformations  $S$ that, at each point $X$  of the spacetime, belong to the spin group ${\sf Spin(1,3)}$ [a relevant definition of it is given after Eq. (\ref{S=S(L)}) below]. Unitary matrices are generally not elements of ${\sf Spin(1,3)}$. The DFW theory is therefore incomplete for quantum mechanics in some aspects by requiring separate treatment for the unitarily equivalent Dirac representation, chiral representation, and the Majorana representation, none of which are related by the spin group ${\sf Spin(1,3)}$. We shall base our discussion of the DFW version of the Dirac equation on the following three properties, which characterize the DFW theory as it is used in most of the literature:
\\

a)	The Dirac matrices  $\gamma ^\mu $  for DFW are of the form  $\gamma ^\mu=a^\mu_{\ \,\alpha}  \ \gamma ^{ \natural \alpha}$, where $a^\mu_{\ \,\alpha} $  are real coefficients such that $a^\mu _{\ \,\alpha} \,a^\nu _{\ \,\beta } \,\eta ^{\alpha \beta }=g^{\mu \nu }$, and where $\gamma ^{ \natural \alpha}$ are a {\it fixed choice} of constant Dirac matrices, taken from a standard representation ({\it i.e.,} chosen to be in the Dirac representation, or in the chiral representation, or in the Majorana representation, or in any other representation that is {\it unitarily} related to these standard representations.) \\

b)	The hermitizing matrix for the Dirac matrices $\gamma ^\mu $  is $A=\gamma ^{\natural 0} $.  This choice of constant hermitizing matrix is invariant under all {\it local spin transformations} $S$.\\

c)	The coefficient fields $(\gamma ^\mu,A) $  for DFW are {\it covariantly constant} with respect to the spin connection of the form: $\Gamma _\nu =\omega _{\nu \rho \tau }s^{\rho \tau }$  with real coefficients $\omega _{\nu \rho \tau }$  and $s^{\rho \tau }\equiv [ \gamma ^{ \natural \rho },\gamma ^{ \natural \tau }]$. That is, all covariant derivatives of the coefficient fields  $(\gamma ^\mu,A) $ for DFW vanish.\\

\vspace{7mm} In a previous work \cite{A42}, it was proved that, in order that any solution $\psi $ of the Dirac equation (\ref{Dirac-general}) satisfy the current conservation
\be\label{current-conservation}
D_\mu J^\mu=0,
\ee 
it is necessary and sufficient that
\be\label{D_mu(B_mu)=0}
D_\mu B^\mu=0\qquad (B^\mu \equiv A\gamma ^\mu).
\ee
Therefore, not all possible coefficient fields $(\gamma ^\mu ,A)$ are physically admissible for the ``normal" Dirac equation (\ref{Dirac-general}), but merely the ones which, in addition to the anticommutation relation (\ref{Clifford}), satisfy condition (\ref{D_mu(B_mu)=0}). Such coefficient fields we call {\it admissible} for the ``normal" Dirac equation (\ref{Dirac-general}). However, it was also proved \cite{A42} that the current conservation (\ref{current-conservation}) applies, for {\it any} coefficient fields, to any $\psi $ solution of the following {\it ``modified Dirac equation:"} 
\be\label{Dirac-general-modified}
\gamma ^\mu D_\mu\psi=-im\psi-\frac{1}{2}A^{-1}(D_\mu B^\mu )\psi.\\
\ee
In other words: for the latter equation, all coefficient fields $(\gamma ^\mu ,A)$, whose $\gamma ^\mu$ fields obey (\ref{Clifford}), are admissible. Note that {\it the modified Dirac equation (\ref{Dirac-general-modified}) coincides with the normal one (\ref{Dirac-general}) for all $\psi $, iff the condition (\ref{D_mu(B_mu)=0}) is satisfied.} In particular, for DFW, the hermitizing matrix field is given by Eq. (\ref{A= A^natural}). Then, since the coefficient fields $(\gamma ^\mu ,A)$ are covariantly constant, the condition (\ref{D_mu(B_mu)=0}) is automatically satisfied \cite{A42}. Thus, {\it for DFW, the modified Dirac equation (\ref{Dirac-general-modified}) coincides with the normal one (\ref{Dirac-general})}.\\

However, for {\it TRD} theory, the condition for current conservation (\ref{D_mu(B_mu)=0}) represents a nontrivial linear constraint of sixteen scalar PDE's to be satisfied by the coefficient fields $(\gamma ^\mu ,A)$. By considering as the unknown the matrix $S$ of a similarity (\ref{similarity-gamma}
)--(\ref{similarity-A}), one may go from any possible coefficients fields $(\gamma ^\mu ,A)$ to ones $(\widetilde {\gamma }^\mu,\widetilde {A})$ which are admissible for the ``normal" Dirac equation (\ref{Dirac-general}), by solving a linear system of sixteen scalar PDE's for the sixteen unknowns of $S$ \cite{A42}. But, in practice, this would involve heavy numerical calculations. Even in the spherical static situation with diagonal, spatially-isotropic metric: $(g_{\mu \nu })=\mathrm{diag}(d_\mu)$ with $d_0=\varphi ^{-1},\quad d_1=d_2=d_3=-\varphi $, $\ \varphi =\varphi (r),\ r\equiv \sqrt{(x^1)^2+(x^2)^2+(x^3)^2}$, we found that the condition (\ref{D_mu(B_mu)=0}) is not satisfied by the most obvious choice of Dirac matrices. 
Namely, for TRD, while there exist fields of Dirac matrices $\gamma ^\mu$  that satisfy the condition (\ref{D_mu(B_mu)=0}), they are generally not so readily obtained as in Eq. (\ref{flat-deformed}).
\\

Therefore, to use the TRD theory, we must have recourse to the modified Dirac equation (\ref{Dirac-general-modified}) with automatic current conservation. At the same time, switching from the normal Dirac equation (\ref{Dirac-general}) to the modified one (\ref{Dirac-general-modified}) is entirely vacuous for DFW, since, as stated above, (\ref{Dirac-general}) and (\ref{Dirac-general-modified}) coincide for DFW. 

\subsection{The Lagrangian and the spin group}\label{Lagrangian}

The modified Dirac equation (\ref{Dirac-general-modified}) is derived from a Lagrangian which generalizes the well known ``Dirac Lagrangian" valid for the DFW equation (e.g. \cite{BrillWheeler1957+Corr,Leclerc2006}). The Lagrangian density is given by:
\be\label{Lagrangian-density}
l=\sqrt{-g}\ \frac{i}{2}\left[\,\overline{\psi}\gamma^{\mu}(D_{\mu}\psi)-
\left(\overline{D_{\mu}\psi} \right)\gamma^{\mu}\psi+2im\overline{\psi}\psi\right],
\ee
where $\overline{\psi}\equiv \psi ^\dagger A$  is the generalized adjoint of $\psi $ and similarly $\overline{D_{\mu}\psi}\equiv \left(D_{\mu}\psi \right) ^\dagger A$. In the general case that we are considering, it is straightforward to check that the Euler-Lagrange equations of this Lagrangian give the generalized Dirac equation (\ref{Dirac-general-modified}).  \\

The point dependent gauge transformations that leave the Lagrangian invariant form a transformation group, say $\mathrm{G_{gauge}}$. This transformation group can equivalently be defined as the set of those local similarity transformations under which the Dirac equation is covariant. For the DFW theory, the Dirac matrices are defined by Eq. (\ref{flat-deformed}). Then the group $\mathrm{G_{gauge}}$ is made of those (smooth) local similarities which are deduced from a change of the tetrad field $u_\alpha =a^\mu _{\ \, \alpha }\,\partial _\mu$, entering Eq. (\ref{flat-deformed}), by a (proper) local Lorentz transform $L=L(X) \in \mathsf{SO(1,3)}$. This is well known, see e.g. on p. 468 of Ref. \cite{BrillWheeler1957+Corr}. Then one may show \cite{A42} that we have Eq. (\ref{similarity-gamma}) with
\be\label{S=S(L)}
S(X)\equiv \pm \mathsf{S}(L(X)),
\ee
where $L\mapsto \pm \mathsf{S}(L)$ is the spinor representation (defined only up to a sign). The set of all $\mathsf{S}(L)$ for $L \in \mathsf{SO(1,3)}$ makes the spin group $\mathrm{G}=\mathsf{Spin(1,3)}$. Thus,  $\mathrm{G_{gauge}}$ for DFW is made of all (smooth) local similarities which at any point belong to the spin group $\mathsf{Spin(1,3)}$. On the other hand, for the  {\it TRD theory,} the transformation group $\mathrm{G_{gauge}}$ was described by a partial differential equation in Eq. (69) of Ref. \cite{A42}. \\

In Subsect. \ref{H-Equivalence-Definition} we consider a possibly larger group of local similarity transformations which preserve the form of each theory, but are not covariant transformations of that theory.

\subsection{Hamiltonian of the modified Dirac equation and Fundamental Formula}\label{Hamiltonian-modified}

Multiplying the modified Dirac equation (\ref{Dirac-general-modified}) by $i\gamma ^0$ on the left, using the anticommutation relation (\ref{Clifford}) and the definition (\ref{alpha}) of the alpha matrices, the Hamiltonian operator H$'$ associated with the modified Dirac equation (\ref{Dirac-general-modified}), analogous to the Hamiltonian operator H in Eq. (\ref{Hamilton-Dirac-normal}), is given by:
\be \label{H-Dirac-modified-brut}
 \mathrm{H}' \equiv  m\alpha  ^0 -i\alpha ^j D _j -i\Gamma_0-\frac{i}{2}\frac{\gamma ^0}{g^{00}}A^{-1}(D_\mu B^\mu ).
\ee
Using again (\ref{Clifford}), the inverse of the hermitizing matrix $B\equiv B^0\equiv A\gamma ^0\ $ is seen to be 
\be\label{B^{-1}}
B^{-1}=\frac{\gamma ^0}{g^{00}}A^{-1},
\ee
hence we may rewrite
\be\label{Hamilton-Dirac-modified}
 \mathrm{H}' \equiv  m\alpha  ^0 -i\alpha ^j D _j -i\Gamma_0-\frac{i}{2}B^{-1}(D_\mu B^\mu )= \mathrm{H}+iC, \qquad C\equiv -\frac{1}{2}B^{-1}(D_\mu B^\mu ).  
\ee
We will now identify the relevant Hilbert space scalar product between wave functions, say $(\psi\mid \varphi )$, and derive the hermiticity condition of $\mathrm{H}'$ for that scalar product. Due to Eq. (\ref{Hamilton-Dirac-modified}), these two steps are only a slight modification of the corresponding steps \cite{A42} for the Hamiltonian $\mathrm{H}$ of the ``normal" Dirac equation, hence we shall be brief. To ensure compatibility with the unique scalar product identified for affine coordinates in a flat spacetime \cite{A40}, the scalar product should have the general form
\footnote{\ 
The second equality assumes that the ``space" manifold M (see the end of Subsect. \ref{H-normal}) is diffeomorphic to ${\sf R}^3$. This is by no means an essential assumption.
}
\be \label{Hermitian-general-sqrt-g}
(\psi  \mid \varphi  ) \equiv \int_\mathrm{M} \ (\psi (x) : \varphi(x))\  \dd {\sf V}(x)=\int_{\sf R^3} \psi^\dagger B  \varphi \ \sigma \ \sqrt{-g}\ \dd^ 3{\bf x}, 
\ee
where $(u : v )$ is a positive-definite Hermitian product between 4-vectors (for TRD) or quadruplets of scalar fields (for DFW), defined by \cite{A40,A42}:
\be\label{B-product}
(u : v )\equiv (u : v )_B \equiv B_{\rho \nu } u ^{\rho*} v^\nu = u^\dagger B v, \qquad B\equiv A\gamma ^0,
\ee
and where $\sigma (t,x)$  is a priori any spatial scalar field [{\it i.e.,} scalar under transformations (\ref{purely-spatial-change})]. From (\ref{Hamilton-Dirac-modified}) and (\ref{B-product}), we find that 
\be\label{delta-hermit}
(\psi : \mathrm{H}'\psi )-(\mathrm{H}'\psi :\psi ) =(\psi : \mathrm{H}\psi )-(\mathrm{H}\psi :\psi )+i[(\psi : C\psi )+(C\psi :\psi )],  
\ee
with
\be\label{Cpsi}
(\psi : C\psi )=-\frac{1}{2}\psi ^\dagger B B^{-1}(D_\mu B^\mu )\psi =-\frac{1}{2}\psi ^\dagger (D_\mu B^\mu )\psi=(C\psi :\psi ).
\ee
Therefore, using Eq. (92) in Ref. \cite{A42}, which applies to a {\it time-independent} $\psi $:
\be \label{delta-hermit-time-independent}
(\psi : \mathrm{H}\psi )-(\mathrm{H}\psi :\psi )=-i[\psi ^\dagger A \gamma^\mu  D_\mu \psi +( D_\mu \psi ) ^\dagger A\gamma ^\mu \psi],
\ee
we get for a such $\psi $:
\bea\label{delta-hermit-2}\nonumber
(\psi : \mathrm{H}'\psi )-(\mathrm{H}'\psi :\psi ) & = & -i\left[\psi ^\dagger A \gamma^\mu  D_\mu \psi +( D_\mu \psi ) ^\dagger A\gamma ^\mu \psi+\psi ^\dagger (D_\mu B^\mu )\psi \right ]\\ \nonumber
& = & -iD_\mu (\psi ^\dagger A \gamma^\mu  \psi) \\ 
& \equiv & -iD_\mu J^\mu.
\eea
This is Eq. (93) of Ref. \cite{A42}, which was derived there for the normal Dirac Hamiltonian (\ref{Hamilton-Dirac-normal}), by assuming then the condition for current conservation (\ref{D_mu(B_mu)=0}). Thus, it is true for the modified Dirac equation, {\it without} assuming the validity of (\ref{D_mu(B_mu)=0}). By integrating (\ref{delta-hermit-2}) in space, using the general form (\ref{Hermitian-general-sqrt-g}) of the scalar product, we derive first in exactly the same way as in Ref. \cite{A42}:
\be\label{delta-hermit-admissible}
i[(\psi \mid  \mathrm{H}'\psi )-(\mathrm{H}'\psi \mid \psi )]= \int_{{\sf R}^3} \psi ^\dagger \left[ \sigma\ \partial _0 \left(\sqrt{-g}\,  A \gamma^0  \right)-(\partial _j\sigma)\,\sqrt{-g}\, A \gamma^j  \right] \psi \, \dd ^3{\bf x}.
\ee
From this, follow the same consequences, now for the modified Dirac equation (\ref{Dirac-general-modified}), as the ones derived in Ref. \cite{A42} for the normal Dirac equation (\ref{Dirac-general}). Namely:\\

\noindent {\bf i}) {\it Scalar product.} (Theorem 5 of Ref. \cite{A42}.) In order that the following two axioms be simultaneously valid:

\paragraph{Axiom (A):}\label{AxiomA}  {\it The Hilbert space scalar product (\ref{Hermitian-general-sqrt-g}) of any time-independent wave functions $\psi $ and $\varphi $ defined on the space manifold $\mathrm{M}$ is time independent,}

\paragraph{Axiom (B):}\label{AxiomB}  {\it  For each time $t$, the Hamiltonian  $\mathrm{H}'$ is a Hermitian ({\it i.e.,} symmetric) operator with respect to the scalar product (\ref{Hermitian-general-sqrt-g}),}\\

\noindent it is {\it necessary} that the scalar field $\sigma $ be a constant, and thus may be assumed to be $\sigma \equiv 1$, so that the scalar product must be

\be \label{Hermitian-sigma=1-g}
(\psi  \mid \varphi  ) \equiv \int_\mathrm{M} (\psi :\varphi )_B \ \dd {\sf V}=\int \psi^\dagger B \varphi\ \sqrt{-g}\ \dd^ 3{\bf x},\qquad \dd {\sf V}\equiv \sqrt{-g}\ \dd^ 3{\bf x}.
\ee
As noted in Ref. \cite{A42}, Axioms (A) and (B) are a somewhat weaker version of two axioms valid for standard quantum mechanics in a flat spacetime.\\

\vspace{6mm}
\noindent {\bf ii}) {\it Hermiticity condition.} (Theorem 6 of Ref. \cite{A42}, now without the necessity to fulfil the $D_\mu B^\mu=0$ condition.) In order that the modified Hamiltonian (\ref{Hamilton-Dirac-modified}) be Hermitian (at time $t$) for the scalar product (\ref{Hermitian-sigma=1-g}), it is necessary and sufficient that
\be\label{hermiticity-condition-}
\partial _0 M({\bf x}) =0 \quad \mathrm{for\ every\ } {\bf x} \in {\sf R}^3, \qquad M\equiv \sqrt{-g}\ B.  \
\ee

\vspace{4mm}
\noindent Moreover, since the scalar product must be (\ref{Hermitian-sigma=1-g}), {\it i.e.,} $\sigma \equiv 1$, Eq. (\ref{delta-hermit-admissible}) becomes the following {\it Fundamental Formula:}
\be\label{FundamentalFormula}
i[(\psi \mid  \mathrm{H}'\psi )-(\mathrm{H}'\psi \mid \psi )]= \int_{{\sf R}^3} \psi ^\dagger \ (\partial _0 M)\ \psi \ \dd ^3{\bf x}.
\ee
This formula, as well as the condition $\sigma \equiv 1$ of Eq. (\ref{delta-hermit-admissible}), applies of course also to the ``normal" Dirac Hamiltonian H in Eq. (\ref{Hamilton-Dirac-normal}) (\cite{A42}, Eq. (96) with $\sigma \equiv 1$).

\vspace{4mm}
\section{Equivalence condition of the Hamiltonians corresponding with two coefficient fields}\label{EquivalentHamiltonians}
\subsection{Definition of equivalent operators}\label{H-Equivalence-Definition}

As was noted in the \hyperref[flat-deformed]{Introduction}, there is an infinity of possible choices for the coefficient fields of the Dirac equation in a given spacetime $(\mathrm{V},g_{\mu \nu })$. As discussed in Subsect. \ref{modified Dirac equation}, if $(\gamma ^\mu,A)$ and $(\widetilde {\gamma}^\mu, \widetilde {A})$ are two possible choices for the coefficient fields, we assume that they are related together by a local similarity $S(X)$ as in Eqs. (\ref{similarity-gamma}) and (\ref{similarity-A}). (See Footnote 3 there.)  We investigate the changes in the Dirac Hamiltonian  (\ref{Hamilton-Dirac-modified}), in the energy operator 
(\ref{H^s = E}) below, and in its spectrum, that follow from a change of the coefficient fields $(\gamma ^\mu,A)$ through a local similarity
(\ref{similarity-gamma})--(\ref{similarity-A}). When doing so, the local similarity transformations do not necessarily leave the Dirac equation covariant under the transformation (\ref{psitilde=S^-1 psi}) of the wave function. Indeed, while the Dirac equations obtained from different choices of the coefficient fields are not all equivalent as wave equations, each of them is derived from a Lagrangian and has a conserved probability current. [For the modified Dirac equation (\ref{Dirac-general-modified}), any solution of the Dirac equation with the new coefficients satisfies the current conservation (\ref{current-conservation}), see Subsect. \ref{modified Dirac equation}
.] While the admissible local similarities $S$ ensure that the anticommutation relation (\ref{Clifford}) holds true, they must also be consistent with the form assumed for the coefficient fields. Recall the three properties (a), (b), (c) of the DFW theory discussed in the paragraph immediately following Eq. (\ref{J-tilde}). \\

Thus, for the {\it TRD theory,} any local similarity $S$ which at any point belongs to ${\sf GL(4,C)}$ is admissible in this generalized sense [with the modified Dirac equation (\ref{Dirac-general-modified})]. However, for the {\it DFW theory,} one assumes the particular form (\ref{flat-deformed}) for the field $\gamma ^\mu $. As recalled in Subsect. \ref{Lagrangian}, this implies that the admissible local similarities belong at any point to the spin group $\mathsf{Spin(1,3)}$, thus any of them leaves the Lagrangian (\ref{Lagrangian-density}) invariant. \\

With each of the two coefficient fields: $(\gamma ^\mu ,A)$ and $(\widetilde {\gamma}^\mu,\widetilde {A})$, corresponds a unique scalar product: (\ref{Hermitian-sigma=1-g}) for the first one, and
\be \label{Hermitian-tilde}
(\chi  \,\widetilde { \mid} \,\omega   ) \equiv \int (\chi :\omega )_{\widetilde{B}}\, \dd {\sf V} =\int \chi ^\dagger \,\widetilde {B} \,\omega \ \sqrt{-g}\,\dd^ 3{\bf x},\qquad \widetilde {B} \equiv  \widetilde {A}\,\widetilde {\gamma}^0
\ee
for the second one. These two scalar products are {\it isometrically equivalent} through the transformation (\ref{psitilde=S^-1 psi}) of the wave function: from (\ref{similarity-gamma}), (\ref{similarity-A}) and (\ref{Hermitian-tilde})$_2$, we get 
\be\label{Btilde = S^croix B S}
\widetilde {B} \equiv S^\dagger B S, 
\ee
whence
\be \label{tilde=isometry}
(\widetilde { \psi }  \,\widetilde { \mid} \,\widetilde { \varphi }  ) \equiv \int_{{\sf R}^3} (S^{-1}\psi ) ^\dagger \,S^\dagger B S \,(S^{-1}\varphi  ) \sqrt{-g}\, \dd^ 3{\bf x}=(\psi \mid \varphi ).
\ee
The Dirac Hamiltonian is H$'$ [Eq. (\ref{Hamilton-Dirac-modified})] for the coefficient fields $(\gamma ^\mu ,A)$, and its rewriting with tildes for $(\widetilde {\gamma}^\mu,\widetilde {A})$:
\be\label{H'tilde}
 \widetilde {\mathrm{H}}' \equiv  mS^{-1}\alpha  ^0 S -iS^{-1}\alpha ^j S \widetilde {D} _j -i\widetilde {\Gamma}_0-\frac{i}{2}\widetilde {B}^{-1}(\widetilde {D}_\mu \widetilde {B}^\mu ).
\ee
For TRD, we have $\widetilde {D} _\mu =D _\mu$ and $\widetilde {\Gamma}_\mu =\Gamma_\mu $. However, for DFW, the covariant derivatives $D_\mu $ and the associated connection matrices $\Gamma _\mu $ depend on the fields $\gamma^\mu $, hence are denoted by $\widetilde {D} _\mu $ and $\widetilde {\Gamma}_\mu $ for $(\widetilde {\gamma}^\mu,\widetilde {A})$. Recall also that, {\it for DFW,} the last term in Eq. (\ref{H'tilde}) vanishes, so that the prime may be omitted. The general reasoning below applies to any quantum-mechanical operator, thus it applies to H and to H$'$ as well; so we shall omit the prime also for TRD. \\

The Hamiltonian operator H is acting on a space of functions (or ``states") $\psi $, each of which is defined on the ``space" manifold M that is associated with the reference frame considered (see the end of Subsect. \ref{H-normal}). The set $\mathcal{H}$ of all wave functions $\psi$ defined on $\mathrm{M}$ that are square-integrable with respect to the scalar product (\ref{Hermitian-sigma=1-g}) form a Hilbert space $\mathcal{H}$. As we know, the operator H acts on a dense subspace $\mathcal{D}$ of $\mathcal{H}$, the domain of H. We shall not need a precise definition of this subspace of $\mathcal{H}$ for the operator (\ref{Hamilton-Dirac-modified}), but only the fact that it is dense. Being an operator with dense domain $\mathcal{D}$, H is fully determined by the set of the products $(\mathrm{H} \, \psi \mid    \varphi  ), \psi ,\varphi \in \mathcal{D}$. 
\footnote{\ 
Indeed, fix any $\psi \in \mathcal{D}$, and set $\theta \equiv \mathrm{H} \, \psi \in \mathcal{H}$. The linear form $\varphi \mapsto F(\varphi )\equiv (\theta \mid \varphi )$ is continuous on $\mathcal{H}$. Since $\mathcal{D}$ is dense in $\mathcal{H}$, this linear form, hence the vector $\theta \in \mathcal{H}$, are determined by the values $F(\varphi )$ for $\varphi \in \mathcal{D}$, thus by the products $(\mathrm{H} \, \psi \mid    \varphi  ), \varphi \in \mathcal{D}$.
}
In particular, these products determine the spectrum of the operator H. [The spectrum of H is real when H is Hermitian (more precisely self-adjoint). When that is not the case, the energy operator E should be substituted for the Hamiltonian operator H in all allusions to the spectrum or to the stationary states: see the beginning of Section \ref{Equivalent-E-Operators} for the definition of E and its relation with H.] In the same way, of course, the operator $\widetilde {\mathrm{H}}$, corresponding just to another possible choice for the coefficient fields, is fully determined by the set of products $(\widetilde{\mathrm{H} } \,\chi  \ \widetilde {\mid }\   \omega   ), \chi ,\omega  \in\ \widetilde{\mathcal{D}}$, with $\widetilde{\mathcal{D}}$ the dense domain of $\widetilde {\mathrm{H}}$ in the Hilbert space $\widetilde{\mathcal{H}}$, endowed with the scalar product (\ref{Hermitian-tilde}). Therefore, from the isometry (\ref{tilde=isometry}) of $\mathcal{H}$ onto $\widetilde{\mathcal{H}}$: $\psi \mapsto \widetilde {\psi }$, it follows that H and $\widetilde {\mathrm{H}}$ are mathematically and physically {\it equivalent} iff, for all $\psi ,\varphi $ in the domain $\mathcal{D}$ of H, we have
\be \label{Hamiltonian-matrix-conserved}
(\widetilde{\mathrm{H} }  \,\widetilde {\psi} \ \widetilde {\mid }\   \widetilde {\varphi}  )=(\mathrm{H} \, \psi \mid    \varphi  ). 
\ee
In particular, the operators H and $\widetilde {\mathrm{H}}$ have then the same spectrum. However, we get directly from (\ref{tilde=isometry}):
\be\label{(H-psi)-tilde!phitilde}
(\widetilde{\mathrm{H}  \,\psi} \ \widetilde {\mid }\   \widetilde {\varphi}  )=(\mathrm{H} \, \psi \mid    \varphi  ). 
\ee
Hence, in order that (\ref{Hamiltonian-matrix-conserved}) be true, it is necessary and sufficient that, for all $\psi \in \mathcal{D}$, we have 
\be
\widetilde{\mathrm{H}  \,\psi}=\widetilde{\mathrm{H} }  \,\widetilde {\psi},
\ee
and since, from (\ref{psitilde=S^-1 psi}), we have $\widetilde{\mathrm{H}  \,\psi}\equiv S^{-1}\,\mathrm{H}\psi \equiv S^{-1}\,\mathrm{H}S\widetilde {\psi}$, this rewrites as
\be \label{similarity-invariance-H}
\widetilde{\mathrm{H} }  =  S^{-1}\,\mathrm{H}\, S.
\ee
This is the condition of equivalence of the Dirac Hamiltonians associated with two different choices of the coefficient fields. This notion of equivalence is physically relevant to the problems related with the spectrum and the stationary states. Note that stationary solutions in the standard sense exist for time-independent coefficient fields in a stationary metric. In the general time-dependent case, in particular for a non-stationary metric $g_{\mu \nu }$, at each given time $t_0$, one may ``freeze" the Hamiltonian operator H$_{t_0}$ of that time, and search for stationary solutions of the corresponding Schr\"odinger equation. One thus obtains, at each time $t_0$, an energy spectrum and a set $\mathcal{S}_{t_0}$ of ``stationary states". The significance of stationary states, even for a time-varying Hamiltonian operator H$_{t_0}$, is that according to quantum mechanics:  if a particle is in a state $\psi $, then an energy measurement at time $t$ will yield an eigenvalue of H$_{t_0}$.  Immediately following the measurement, the state $\psi $ will collapse into an eigenvector of H$_{t_0}$, as a result of the measurement.

\subsection{Equivalence condition of the Hamiltonians: DFW}\label{H-Equivalence-DFW}
Let us investigate when exactly does a change of the coefficient fields lead to an equivalent Hamiltonian, {\it i.e.,} when do we have Eq. (\ref {similarity-invariance-H})? For the standard version of the Dirac equation in a curved spacetime, which we denote by DFW, we may consider the ``normal" Dirac equation (\ref{Dirac-general}), hence the ``normal" Dirac Hamiltonian (\ref{Hamilton-Dirac-normal}), as noted at the end of Subsect. \ref{modified Dirac equation}. For DFW, the connection matrices change after the similarity transformation (\ref{similarity-gamma}) in the following way \cite{ChapmanLeiter1976}:
\be\label{Gamma-tilde-psitilde=S^-1 psi}
\widetilde{\Gamma }_\mu = S^{-1}\Gamma _\mu S+ S^{-1}(\partial _\mu S).
\ee
For both DFW and TRD, this is in fact the {\it characteristic} relation to be satisfied by the $\widetilde{\Gamma }_\mu$ 's, in order that the Dirac equation (\ref{Dirac-general}) be covariant under the similarity (\ref{similarity-gamma})--(\ref{psitilde=S^-1 psi})---as one easily checks. For DFW, it is always satisfied when $S$ has the required form (\ref{S=S(L)}). For TRD, the local similarity $S$ is arbitrary, but the $\Gamma _\mu$ 's are {\it invariant,} so that the Dirac equation (\ref{Dirac-general}) generally is {\it not} covariant under the similarity (\ref{similarity-gamma})--(\ref{psitilde=S^-1 psi}) \cite{A42}. From (\ref{Hamilton-Dirac-normal}) with the tilded fields, and from (\ref{Gamma-tilde-psitilde=S^-1 psi}), we get for DFW:
\bea\nonumber
 \widetilde {\mathrm{H}} & = & mS^{-1}\alpha  ^0 S -iS^{-1}\alpha ^j S (\partial _j + S^{-1}\Gamma _j \,S + S^{-1}(\partial _j S)) -iS^{-1}\Gamma_0 S -iS^{-1}(\partial _0 S),
\eea
while we get directly from (\ref{Hamilton-Dirac-normal}):
\be
 S^{-1}\mathrm{H}S =  mS^{-1}\alpha  ^0 S -iS^{-1}\alpha ^j (\partial _j + \Gamma _j )S -iS^{-1}\Gamma_0 S.
\ee
Moreover, we have by the composition rule of operators and Leibniz' rule:
\be
  S \partial  _j = \partial _j S-(\partial _j S),\qquad (\partial _j S): \psi \mapsto (\partial _j S)\psi,
\ee
whence 
\bea\label{Htilde-S^-1 H S}\nonumber
\widetilde {\mathrm{H}}- S^{-1}\mathrm{H}S & = & -iS^{-1}\alpha ^j S \partial _j -iS^{-1}\alpha ^j(\partial _j S)-iS^{-1}(\partial _0 S)+ iS^{-1}\alpha ^j \partial _j S \\
& = & -iS^{-1}(\partial _0 S).
\eea
Therefore, the DFW Hamiltonian operator $\widetilde {\mathrm{H}}$ corresponding with the new (tilded) coefficient fields is physically equivalent to the one, H, corresponding with the starting coefficient fields ($\widetilde {\mathrm{H}}= S^{-1}\mathrm{H}S$), iff
\be\label{partial_0 S=0}
\partial_0 S=0,
\ee
that is, iff the similarity $S$ is independent of the time $t\equiv x^0$---in the reference frame F which is considered, see the end of Subsect. \ref{H-normal}. Note that, in the derivation of the condition for equivalence, Eq. (\ref{partial_0 S=0}), we used only the easily checked expression (\ref{Hamilton-Dirac-normal}) of the DFW Hamiltonian, together with the standard transformation (\ref{Gamma-tilde-psitilde=S^-1 psi}) of the spin connection.\\

However, nothing prevents one from changing the coefficients by a {\it time-dependent} similarity, which leads to inequivalent Hamiltonians. {\it Thus, the DFW Hamiltonian operator is non-unique.} If the metric $g_{\mu \nu }$ is time-independent (in the frame F), we may choose coefficient fields $(\gamma ^\mu ,A)$ that are independent of $t$. However, in the general case ($g_{\mu \nu ,0} \ne 0$), any possible coefficient fields will depend on $t$. In that case, there seems to be no way to define a unique Hamiltonian operator. \\

\subsection{Equivalence condition of the Hamiltonians: TRD}\label{H-Equivalence-TRD}

For TRD, the Hamiltonian with the new coefficients, Eq. (\ref{H'tilde}), is
\be\label{H'tilde-TRD}
 \widetilde {\mathrm{H}}' \equiv  mS^{-1}\alpha  ^0 S -iS^{-1}\alpha ^j S D _j -i\Gamma_0-\frac{i}{2}\widetilde {B}^{-1}(D_\mu \widetilde {B}^\mu ). 
\ee
In TRD, the components $S^\mu _{\ \,\nu} $ of the similarity matrix $S$ build a $(^1 _1)$ tensor, whose covariant derivative may be written in matrix form as 
\be\label{D_mu S}
D_\mu S \equiv \partial _\mu S+\Gamma _\mu S-S\Gamma _\mu.
\ee
We have then by the composition rule of operators and Leibniz' rule:
\be
  S D _j = D_j S-(D_j S),\qquad (D_j S): \psi \mapsto (D_j S)\psi,
\ee
and also, since, from (\ref{similarity-gamma}) and (\ref{similarity-A}), $\widetilde {B}^\mu =S^\dagger B^\mu S$ and $\widetilde {B}=\widetilde {B}^0$:
\bea
 \widetilde {B}^{-1}(D_\mu \widetilde {B}^\mu )& = & S^{-1}B^{-1}(S^\dagger )^{-1}[S^\dagger  (D_\mu B^\mu )S+(D_\mu S)^\dagger  B^\mu S + S^\dagger  B^\mu (D_\mu S)]\\\nonumber
 & = & S^{-1}[B^{-1} (D_\mu B^\mu )]S+ S^{-1}B^{-1}[B^\mu (D_\mu S) + (S^\dagger)^{-1} (D_\mu S)^\dagger  B^\mu  S].
\eea
So that [using the fact that $B^{-1}B^0={\bf 1}$ and that, from Eq. (\ref{B^{-1}}) above, $B^{-1}B^j=\alpha ^j$]:
\bea\nonumber\label{H'tilde-TRD--2}
 \widetilde {\mathrm{H}}' & =  & mS^{-1}\alpha  ^0 S -iS^{-1}\alpha ^j (D_j S-(D_j S)) -i\Gamma_0-\frac{i}{2}S^{-1}[B^{-1} (D_\mu B^\mu )]S \\
& & -\frac{i}{2}S^{-1}[(D_0S)+\alpha ^j(D_j S)+ B^{-1} (S^\dagger)^{-1} (D_\mu S)^\dagger  B^\mu  S]. 
\eea
On the other hand, we have directly from (\ref{Hamilton-Dirac-modified}):
\be\label{S^{-1}H'S-TRD}
 S^{-1}\mathrm{H}'S =  mS^{-1}\alpha  ^0 S -iS^{-1}\alpha ^j D _jS -iS^{-1}\Gamma_0 S-\frac{i}{2}S^{-1}[B^{-1} (D_\mu B^\mu )]S. 
\ee
Therefore [multiplying (\ref{H'tilde-TRD--2}) and (\ref{S^{-1}H'S-TRD}) by $S$ on the left], $\widetilde{\mathrm{H}}'=S^{-1}\mathrm{H}'S\ $ is equivalent to
\bea
i\alpha ^j (D_j S) -iS\Gamma_0-\frac{i}{2}[(D_0S)+\alpha ^j(D_j S)+ B^{-1} (S^\dagger)^{-1} (D_\mu S)^\dagger  B^\mu  S]=-i\Gamma_0S, 
\eea
or [using (\ref{D_mu S})] to
\bea
\frac{1}{2}[\alpha ^j(D_j S)- B^{-1} (S^\dagger)^{-1} (D_\mu S)^\dagger  B^\mu  S]=\frac{1}{2}(D_0S)+S\Gamma_0-\Gamma_0S \equiv (\partial _0S) - \frac{1}{2}(D_0S), 
\eea
which, using the remark made before Eq. (\ref{H'tilde-TRD--2}), may be finally rewritten as
\be\label{H=S^{-1}HS-TRD}
2B(\partial _0S)S^{-1} =B^\mu (D_\mu S)S^{-1} -[B^\mu (D_\mu S)S^{-1}]^\dagger.
\ee
The r.h.s. is antihermitian, but not the l.h.s. These are sixteen, generally independent, quasilinear PDE's, for the sixteen independent unknowns $S^\mu _{\ \,\nu} $. One may choose freely the field of Dirac matrices $\gamma ^\mu$ satisfying the anticommutation (\ref{Clifford}), plus the arbitrary scalar of the corresponding hermitizing matrix $A(X)$, thus getting a starting set of coefficient fields, $(\gamma ^\mu ,A)$, and the corresponding Hamiltonian H$'$, Eq. (\ref{Hamilton-Dirac-modified}). Then, those coefficient fields $(\widetilde {\gamma} ^\mu ,\widetilde {A})$, whose associated Hamiltonian $\widetilde {\mathrm{H}}'$ is equivalent to H$'$, are likely to be determined uniquely, modulo an initial data, by Eq. (\ref{H=S^{-1}HS-TRD}). But the possible coefficient fields are deduced from $(\gamma ^\mu ,A)$ by a {\it fully} arbitrary local similarity $S(X)$, which, most generally, will thus not obey Eq. (\ref{H=S^{-1}HS-TRD}): the vast majority of them are hence {\it not} equivalent to $(\gamma ^\mu ,A)$. In other words, the Hamiltonian of the modified TRD equation is not unique, either. 

\section{Equivalence condition of the energy operators corresponding with two coefficient fields}\label{Equivalent-E-Operators}

The previous section has shown that, for both the standard (DFW) and our alternative (TRD) versions of the gravitational Dirac equation, the Hamiltonian is non-unique: one set of possible coefficient fields $(\gamma ^\mu ,A)$ being given, the vast majority of the other ones $(\widetilde {\gamma} ^\mu ,\widetilde {A})$ lead to a Hamiltonian operator which does not satisfy the condition of equivalence (\ref{similarity-invariance-H}). With this result, the foregoing observation \cite{A42} that, for DFW, the hermiticity of the Hamiltonian is not stable under admissible similarities, is not surprising any more. It also indicates that the spectrum of the Dirac Hamiltonian may be nonunique, too. However, since the Dirac Hamiltonian H (or H$'$ for the modified Dirac equation) is not always Hermitian, the relevant spectrum is rather the spectrum of the {\it Hermitian part} of H or H$'$:
\be
\mathrm{H}^{s}\equiv \frac{1}{2}(\mathrm{H}+\mathrm{H}^\ddagger ),
\ee
where the Hermitian conjugate operator $\mathrm{H}^\ddagger$ is w.r.t. the unique scalar product (\ref{Hermitian-sigma=1-g}). We get from this definition, by using (\ref{FundamentalFormula}) and (\ref{Hermitian-sigma=1-g}):
\bea \label{delta-H-with-F}
(\psi \mid\mathrm{H}^{s}\psi  ) & = & \frac{1}{2}\left[(\psi \mid(\mathrm{H}+\mathrm{H}^\ddagger ) \psi  )\right]\nonumber\\
& = & \frac{1}{2}\left[(\mathrm{H} \psi \mid \psi  )+(\psi \mid \mathrm{H}\psi  )\right]\nonumber\\
& = & (\psi \mid \mathrm{H}\psi  )+\frac{1}{2}\left[(\mathrm{H} \psi \mid \psi  )-(\psi \mid \mathrm{H}\psi  )\right]\nonumber\\
& = & (\psi \mid \mathrm{H}\psi  )+\frac{i}{2}\int \psi^\dagger (\partial _0 M)\psi\, \dd^ 3{\bf x}\qquad (M\equiv \sqrt{-g}\,B)\nonumber\\
& = & (\psi \mid \mathrm{H}\psi  )+\int \psi^\dagger M\left(\frac{i}{2}M^{-1}(\partial _0 M)\right)\psi\, \dd^ 3{\bf x}\nonumber\\
& \equiv & (\psi \mid \mathrm{E}\psi  ),
\eea
where since $M\equiv \sqrt{-g}\,B$,
\be\label{H^s = E}
\mathrm{E} \equiv  \mathrm{H}+\frac{i}{2\sqrt{-g}}\,B^{-1}\,\partial _0 \left(\sqrt{-g}\,B\right) = \mathrm{H}^{s} .
\ee
This result applies to the ``normal" Dirac Hamiltonian H, Eq. (\ref{Hamilton-Dirac-normal}), and to the modified one H$'$ (\ref{Hamilton-Dirac-modified}) as well. This Hermitian-symmetrized operator E or E$'$ coincides, in the particular case envisaged by Leclerc \cite{Leclerc2006}, with what one (including him) calls the {\it energy operator,} which is derived from the field Lagrangian. [The case considered by Leclerc is the DFW equation, with the tetrad matrix $a^\mu _{\ \, \alpha} $ in Eq. (\ref{flat-deformed}) satisfying $a^0 _{\ \, j}=0 $ for $j=1,2,3$.] As with Leclerc, we can show generally that the expected value of the energy operator E$'$ [i.e., the energy operator for the modified Dirac equation] equals the classical field energy as derived from the Lagrangian (\ref{Lagrangian-density}). Of course, the condition of equivalence of the energy operators E and $\widetilde{\mathrm{E}}$ corresponding with two coefficient fields $(\gamma ^\mu ,A)$ and $(\widetilde {\gamma} ^\mu ,\widetilde {A})$ is the same as for the Hamiltonian operators, Eq. (\ref{Hamiltonian-matrix-conserved}) [with E or E$'$ in the place of H]. When it is true, the operators E and $\widetilde{\mathrm{E}}$ have, in particular, the same spectrum. As with H, (\ref{Hamiltonian-matrix-conserved}) is equivalent to (\ref{similarity-invariance-H}), thus here to 
\be \label{similarity-invariance-E}
\widetilde{\mathrm{E} }  =  S^{-1}\,\mathrm{E}\, S.
\ee
In order to see when this is true, we will be able to use the work done in Sect. \ref{EquivalentHamiltonians} for the Hamiltonian, since, from (\ref{Btilde = S^croix B S}) and (\ref{H^s = E}), we get after a straightforward computation:
\be\label{SEtilde-ES}
\widetilde{\mathrm{E}} -S^{-1}\mathrm{E}S = \widetilde{\mathrm{H}} - S^{-1}\mathrm{H}S+\frac{i}{2}\left [S^{-1}B^{-1}(S^\dagger )^{-1}(\partial _0S)^\dagger BS+ S^{-1}(\partial _0S) \right ].
\ee
This is true for the modified Dirac equation as well as for the normal one.

\subsection{Equivalence condition of the energy operators: DFW}\label{E-Equivalence-DFW}

For DFW, we have from (\ref{Htilde-S^-1 H S}) and (\ref{SEtilde-ES}):
\bea\label{SEtilde-ES-DFW}\nonumber
\widetilde{\mathrm{E}} -S^{-1}\mathrm{E}S & = & -iS^{-1}(\partial _0 S)+\frac{i}{2}\left [S^{-1}B^{-1}(S^\dagger )^{-1}(\partial _0S)^\dagger BS+ S^{-1}(\partial _0S) \right ]\\\nonumber
& = & \frac{i}{2}\left [S^{-1}B^{-1}(S^\dagger )^{-1}(\partial _0S)^\dagger BS - S^{-1}(\partial _0S) \right ],\\
S\widetilde{\mathrm{E}}S^{-1} -\mathrm{E} & = & \frac{i}{2}B^{-1}\left\{ \left[B(\partial _0S)S^{-1}\right]^\dagger -B(\partial _0S)S^{-1}\right\}.
\eea
Hence, $\widetilde{\mathrm{E}} = S^{-1}\mathrm{E}S$ occurs exactly when $S$ satisfies
\be\label{SEtilde=ES DFW}
B(\partial _0S)S^{-1}-\left[B(\partial _0S)S^{-1}\right]^\dagger \equiv 2 \left[B(\partial _0S)S^{-1}\right]^a = 0,
\ee
which is the vanishing of an antihermitian matrix, involving sixteen independent real quasilinear equations. Any time-independent similarity is a solution of (\ref{SEtilde=ES DFW}). For an arbitrary allowed similarity, $S(X)={\sf S}(L(X))$, one has in general $\widetilde{\mathrm{E}} \ne S^{-1}\mathrm{E}S$, so that the energy operator of the DFW equation is non-unique. Moreover, the difference operator $S\widetilde{\mathrm{E}}S^{-1} -\mathrm{E}$ is actually a rather general operator, Eq. (\ref{SEtilde-ES-DFW}), hence the spectra of the operators $\mathrm{E}$ and $S\widetilde{\mathrm{E}}S^{-1}$ are likely to be generally different. Section \ref{Non-uniquenessProof} proves that, for any arbitrary coefficient fields $(\gamma ^\mu ,A)$, one may choose the similarity $S$, of the form needed for DFW, so that definitely the spectra of the operators $\mathrm{E}$ and $S\widetilde{\mathrm{E}}S^{-1}$ {\it are} different. But Eq. (\ref{(H-psi)-tilde!phitilde}) with E in the place of H shows that, for any similarity $S$, the spectrum of $\mathrm{E}$ is the same as that of $S^{-1}\mathrm{E}S$, thus symmetrically the spectrum of $S\widetilde{\mathrm{E}}S^{-1}$ is just the same as that of $\widetilde{\mathrm{E}}$. Therefore, the energy spectrum of the DFW equation is non-unique.

\subsection{Equivalence condition of the energy operators: TRD}\label{E-Equivalence-TRD}

For the Hamiltonian of the (modified) TRD equation, we get from Eqs. (\ref{H'tilde-TRD--2}) and (\ref{S^{-1}H'S-TRD}):
\be
\frac{2}{i}\left(S\widetilde{\mathrm{H}}' - \mathrm{H}'S\right)=\alpha ^j(D_j S)- B^{-1} (S^\dagger)^{-1} (D_\mu S)^\dagger  B^\mu  S-(D_0S)-2S\Gamma_0+2\Gamma_0S ,
\ee
hence (\ref{SEtilde-ES}) gives us 
\bea\label{SEtilde-ES-modified}
\frac{2}{i}\left(S\widetilde{\mathrm{E}}' -\mathrm{E}'S \right)& = & \alpha ^j(D_j S)- B^{-1} (S^\dagger)^{-1} (D_\mu S)^\dagger  B^\mu  S-(D_0S)-2S\Gamma_0+2\Gamma_0S \nonumber\\
&  & +B^{-1}(S^\dagger )^{-1}(\partial _0S)^\dagger BS+ (\partial _0S), 
\eea
or, using (\ref{D_mu S}) and the remark made before Eq. (\ref{H'tilde-TRD--2}), 
\be\label{S Etilde S^-1 -E}
B\left(S\widetilde{\mathrm{E}}'S^{-1} -\mathrm{E}' \right )=i \left[B^\mu (D_\mu S)S^{-1}-B(\partial _0S)S^{-1}\right]^a.
\ee
Hence, $\widetilde{\mathrm{E}}' = S^{-1}\mathrm{E}'S$ is equivalent to the vanishing of an antihermitian complex matrix:
\be\label{SEtilde=ES modified}
\left[B^\mu (D_\mu S)S^{-1}-B(\partial _0S)S^{-1}\right]^a =0.
\ee
Thus, the invariance condition of the energy operator of the modified TRD equation imposes only sixteen independent quasilinear real equations (\ref{SEtilde=ES modified}), to the sixteen independent complex components of the similarity $S$. However, not all similarities $S(X)$ (all of which are admissible here) lead to equivalent energy operators. For an arbitrary similarity, again the difference operator $S\widetilde{\mathrm{E}}'S^{-1} -\mathrm{E}'$ is a rather general operator, Eq. (\ref{S Etilde S^-1 -E}), so that the spectra of the operators $\mathrm{E}'$  and $S\widetilde{\mathrm{E}}'S^{-1}$ are generally different. Subsection \ref{nonuniqueness-TRD} shows that $S$ may always be chosen so that they are indeed different. And again the spectrum of $S\widetilde{\mathrm{E}}'S^{-1}$ is the same as that of $\widetilde{\mathrm{E}}'$. Hence, the energy spectrum of the modified TRD equation (with automatic current conservation) is non-unique, too.

\section{Non-uniqueness proof of the Dirac energy spectrum}\label{Non-uniquenessProof}

In this section we prove the following result:
\paragraph{Theorem.}  Let V be any spacetime that is diffeomorphic to ${\sf R}^4$, whose Lorentz metric $g_{\mu \nu }$ satisfies $g_{00}>0$, and whose spatial metric $h_{jk}\equiv -g_{jk}$ is positive definite. Assume $m \ne 0$ in the Dirac equation. Then, for DFW as well as for TRD, for any admissible set of coefficient fields $(\gamma ^\mu,A) $, there exists an admissible local similarity $S$ such that the spectra of the energy operators E and $\widetilde {\mathrm{E}}$ before and after the similarity are different. Consequently, the spectrum of the energy operator for the free Dirac equation is not unique in both DFW and TRD.\\

The proof given below assumes that the matrix $B\equiv A\gamma ^0$ is proportional to the identity. [For DFW, this is true if one assumes that the matrix $(a^\mu _{\ \, \alpha })$ in Eq. (\ref{flat-deformed}) is such that $a^0_{\ \, j}=0$: see after Eq. (\ref{SEtilde-ES-DFW-2}).] As discussed after Eq. (\ref{SEtilde-ES-DFW-2}), the theorem can also be proved without this technical assumption.

\subsection{Perturbation of an eigenstate by a small similarity}

Let us consider a family of local similarities $ S$ depending on a small parameter $\varepsilon $. Thus, $\delta S$ being a possible direction for a perturbation of the starting coefficients by a local similarity close to the identity, we have
\be
S(\varepsilon ,X)=I+\varepsilon \, (\delta S)(X)+O(\varepsilon ^2).
\ee
The local similarity transformation $S(\varepsilon )$ transforms the coefficient fields $(\gamma ^\mu ,A)$ as in Eqs. (\ref{similarity-gamma}) and (\ref{similarity-A}), and thereby transforms the Dirac equation and its associated energy operator E, as well as the Hilbert space scalar product in Eq. (\ref{Hermitian-sigma=1-g}) [see Eq. (\ref{Hermitian-tilde})]. Hence the perturbed energy operator, denoted as $\widetilde {\mathrm{E}}(\varepsilon )$, is generally an operator in a different Hilbert space than the Hilbert space in which the unperturbed energy operator E acts.  It is therefore convenient to define another operator $\mathrm{E}(\varepsilon )$, having the same spectrum as $\widetilde {\mathrm{E}}(\varepsilon )$, by bringing back the perturbed energy operator $\widetilde {\mathrm{E}}(\varepsilon )$ to act in the same Hilbert space as the unperturbed energy operator E as follows:
\be
\mathrm{E}(\varepsilon )=S(\varepsilon )\widetilde{\mathrm{E}}(\varepsilon )S(\varepsilon )^{-1}.
\ee
Let $\psi (\varepsilon )$ be an eigenfunction of the operator $\mathrm{E}(\varepsilon )$, with eigenvalue $\lambda (\varepsilon )$.  There is a corresponding eigenfunction $\widetilde {\psi} (\varepsilon )=S(\varepsilon )^{-1}\psi (\varepsilon )$ of $\widetilde {\mathrm{E}}(\varepsilon )$ with the same eigenvalue $\lambda (\varepsilon )$. Then, using standard perturbation theory of quantum mechanics, we may write:
\be\label{delta lambda}
\delta \lambda \equiv \lambda (\varepsilon )-\lambda (0)= (\psi(0) \mid \delta \mathrm{E}(\varepsilon )\psi(0) )+O(\varepsilon ^2),
\ee
with
\be
\delta \mathrm{E}(\varepsilon )\equiv \mathrm{E}(\varepsilon )-\mathrm{E}(0)=S(\varepsilon )\widetilde{\mathrm{E}}(\varepsilon )S(\varepsilon )^{-1}-\mathrm{E}.
\ee
Note that this result uses only the hermiticity of the unperturbed operator E. 

\subsection{General explicit form of the DFW energy operator}\label{E explicit-DFW}
We have for DFW [Eq. (\ref{SEtilde-ES-DFW})]:
\be\label{SEtilde-ES-DFW-1}
\delta \mathrm{E}\equiv S\widetilde{\mathrm{E}}S^{-1} -\mathrm{E} =-iB^{-1} \left[B(\partial _0S)S^{-1}\right]^a, \qquad Q^a\equiv \frac{1}{2}(Q-Q^\dagger ).
\ee
Recall that, at any spacetime point $X$, we defined a Hermitian product of complex vectors (\ref{B-product}), by using the positive definite Hermitian matrix $B$:
\be\label{B product}
(u:v)_B \equiv u^\dagger B v, \qquad B=B(X).
\ee
It is evident that the adjoint of a matrix $Q$ with respect to that Hermitian product is
\be\label{Q^croix B}
Q^{\dagger (B)} \equiv B^{-1} Q^\dagger B, 
\ee
{\it i.e.,} we have
\be
(Q^{\dagger (B)}u\,:\,v)_B= (u\,:\,Qv)_B.
\ee
Using (\ref{Q^croix B}), we may rewrite (\ref{SEtilde-ES-DFW-1}) as 
\be\label{SEtilde-ES-DFW-2}
\delta \mathrm{E}=-i\left[(\partial _0S)S^{-1}\right]^{a(B)}, \qquad Q^{a(B)} \equiv \frac{1}{2}(Q-Q^{\dagger (B)} ).
\ee
In DFW theory it is permitted to choose any tetrad field, whose matrix   $(a^\mu _{\ \, \alpha })$ satisfies Eq. (\ref{compatible-anticom}). It is generally assumed that one tetrad field is not to be preferred over another. To demonstrate the non-uniqueness of the DFW energy operator it suffices to consider matrices $(a^\mu _{\ \, \alpha })$ such that $a^0_{\ \, j}=0$.\footnote{\ 
The restriction $a^0_{\ \, j}=0$ applies to the starting tetrad field, which defines the starting field of Dirac matrices $\gamma ^\mu $. We do not impose a similar restriction on the new tetrad field, nor equivalently on the similarity $S$, which defines the new matrices $\widetilde {\gamma} ^\mu $.
} 
Such tetrad fields always exist and are almost universally used in the literature.
\footnote{\ 
A version of Cholesky's matrix factorization theorem, which applies to Lorentz metrics $G\equiv (g_{\mu \nu })$ under the assumptions that $g_{00}>0$ and that $h\equiv -(g_{j k })$ is positive definite, states that any matrix $(a^\mu _{\ \, \alpha })$, satisfying Eq. (\ref{compatible-anticom}), factors uniquely as a lower triangular matrix $(t^\mu _{\ \, \alpha })$, with positive entries on the diagonal, also satisfying Eq. (\ref{compatible-anticom}), and a local Lorentz transformation $L ^\beta  _{\ \, \alpha }$. Thus, in DFW, we can set $(a^\mu _{\ \, \alpha })$ equal to a lower triangular matrix (as in fact it is in most of the literature).  
}
Then we get $a^0 _{\ \,0}=\sqrt{g^{00}}$ from the orthonormality of the tetrad, hence:
\be\label{a^0_j=0case}
B\equiv \,A\,\gamma^0 = \sqrt{g^{00}}\, {\bf 1}_4.
\ee
Thus, with the restriction $a^0_{\ \, j}=0$, which we use here for simplicity, either (\ref{SEtilde-ES-DFW-1}) or (\ref{SEtilde-ES-DFW-2}) is simply:
\be\label{SEtilde-ES-DFW-3}
\delta \mathrm{E}=-i\left[(\partial _0S)S^{-1}\right]^{a},
\ee
moreover the product (\ref{B product}) is proportional to the canonical Hermitian product in ${\sf C}^4$, $u.v\equiv u^\dagger v $. However, in the most general case, Eq. (\ref{SEtilde-ES-DFW-2}) allows one to draw the same conclusions as does (\ref{SEtilde-ES-DFW-3}).\\

As is well known: if $t\mapsto S(t)$ is a smooth curve in whatever matrix group G, then $\frac{\dd S}{\dd t}S^{-1}$ belongs to the Lie algebra $\mathcal{G}$ of G, and conversely any element of $\mathcal{G}$ has the form $\frac{\dd S}{\dd t}S^{-1}$ for some smooth curve $t\mapsto S(t) \in \mathrm{G}$.
\footnote{\
Indeed, $\frac{\dd S}{\dd t}$  is a tangent vector at the point $S(t)\in\,$G, hence  $\frac{\dd S}{\dd t}S^{-1}$ is a tangent vector at the identity of G, and thus, by definition, lies in the Lie algebra of G.  Clearly, every infinitesimal generator of G can be derived this way. 
}
For DFW, $S$ has the form $S= \pm \mathsf{S}(L)$, where $L\mapsto \pm \mathsf{S}(L)$ is the spinor representation, defined (up to a sign) for proper Lorentz transformations: $L \in {\sf SO(1,3)}$, Eq. (\ref{S=S(L)}). The spin group $\mathrm{G}=\mathsf{Spin(1,3)}$, set of all such $S$ 's, is known to be isomorphic to ${\sf SL(2,C)}$ and hence, as a real manifold, has dimension six. The explicit form of the spinor representation \cite{Schulten1999} is generated by the commutators of the ``flat" Dirac matrices $\gamma ^{\natural \alpha} $:
\be\label{s^ab}
s^{\alpha \beta } \equiv [\gamma^{\natural \alpha}  ,\gamma^{\natural \beta }  ].
\ee
The six matrices $s^{\alpha \beta  }$, $0\leq \alpha <\beta \leq 3$, are known to be linearly independent (for combinations with real coefficients), thus they form a basis of the real vector space $\mathcal{G}$, the tangent space to G at the identity. Hence, when $t\mapsto S(t)$ is any curve in ${\sf Spin(1,3)}$, the matrix $\frac{\dd S}{\dd t}S^{-1}$ has the form
\be
\frac{\dd S}{\dd t}S^{-1}= \omega_{\alpha \beta} s^{\alpha \beta },
\ee
for some arbitrary real coefficients $\omega_{\alpha \beta }(t)=-\omega_{\beta \alpha }(t)$, determined by the arbitrary curve $S(t)$. It follows from this and from (79) that we have
\be\label{deltaE}
\delta \mathrm{E}=-i \left[\omega_{\alpha \beta} s^{\alpha \beta  }\right]^a,
\ee
where, depending on the local similarity $S(X) \in {\sf Spin(1,3)}$, the six coefficients $\omega_{\alpha \beta}=-\omega_{\beta \alpha }$ can depend arbitrarily on the spacetime point $X$.

\vspace{4mm}
From (\ref{s^ab}) and the definition of the hermitizing matrix $A$, we have in general
\be
(s^{\alpha \beta  })^\dagger =As^{\beta  \alpha }A^{-1}.
\ee
With standard Dirac matrices, for which $A=\gamma ^{\natural 0}=A^{-1}$, we get thus
\be\label{e_ab s^ab ^antisym}
\left[\omega_{\alpha \beta} s^{\alpha \beta  }\right]^a = \sum_{0\leq \alpha <\beta \leq 3} \omega  _{\alpha \beta} \left(s^{\alpha \beta}  -\gamma ^{\natural 0}\, s^{\beta  \alpha }\,\gamma ^{\natural 0} \right).
\ee
Since the anticommutation relation with the Minkowski metric implies that $\gamma ^{\natural 0}\gamma ^{\natural 0}={\bf 1}_4$ and $\gamma ^{\natural 0}\gamma ^{\natural j}=-\gamma ^{\natural j}\gamma ^{\natural 0}$, we have
\be
\gamma ^{\natural 0}\, \gamma ^{\natural \beta } \, \gamma ^{\natural \alpha } \,\gamma ^{\natural 0}=\left\{ \begin{array}{ll} -\gamma ^{\natural \alpha } \,\gamma ^{\natural \beta }  &  \mathrm{if}\ \alpha =j \mathrm{\ and\ }\beta =k ,\quad j,k\in \{1,2,3\},j\ne k,\\
& \\
 +\gamma ^{\natural \alpha } \,\gamma ^{\natural \beta } & \mathrm{otherwise}.
\end{array} \right.
\ee
Hence, (\ref{e_ab s^ab ^antisym}) is in fact
\be\label{e_ab s^ab ^antisym-2}
\left[\omega_{\alpha \beta} s^{\alpha \beta  }\right]^a = 2\sum_{1\leq j <k \leq 3} \omega  _{jk } s^{jk} =\sum_{ j ,k =1}^3 \omega  _{jk } s^{jk}.
\ee

\vspace{4mm}

\subsection{Specialization to the chiral representation}\label{chiral}

For definiteness, we shall now specialize further to the ``chiral" Dirac matrices. However, we mention that, using Eq. (\ref{SEtilde-ES-DFW-2}) and some amount of (Dirac) algebra, the results can be shown to hold independently of the representation, {\it i.e.,} of the set of ``flat" Dirac matrices, and also hold true for any orthonormal tetrad field $a^\mu _{\ \, \alpha }$. If the $\gamma ^{\natural \alpha }$ 's are the ``chiral" matrices, we get \{\cite{Schulten1999}, Eqs. (10.257) and (10.260)\}:
\be\label{e_jk s^jk Chiral}
\sum_{ j ,k =1}^3 \omega _{jk } s^{jk}=-\frac{i}{2}\begin{pmatrix}
\vec {\theta }.\vec {\sigma }  & 0\\ 
 0 &  \vec {\theta }.\vec {\sigma } 
\end{pmatrix},
\ee
where $\vec {\theta }\equiv (\theta _k)$ with $\theta _1\equiv \omega  _{23 }$ (circular), and where $\vec {\sigma }\equiv (\sigma  ^k)$ with $\sigma  ^k$ the Pauli matrices. Putting together Eqs. (\ref{deltaE}), (\ref{e_ab s^ab ^antisym-2}), and (\ref{e_jk s^jk Chiral}), we obtain finally
\be\label{deltaE-final}
\delta \mathrm{E}=P\equiv \begin{pmatrix}
N  & 0\\ 
 0 &  N 
\end{pmatrix}, \qquad N\equiv -\frac{1}{2}\vec {\theta }.\vec {\sigma }.
\ee
The operator $\delta \mathrm{E}$ is thus the multiplication by the point-dependent $4\times 4$ matrix $P$. Depending on $\vec {\theta }$, {\it i.e.,} on the three real numbers $\omega  _{jk}, \ 1\leq j <k \leq 3$, the matrix $N$ can be in fact {\it any} $2\times 2$ Hermitian matrix with zero trace; see {\it e.g.} Ref. \cite{Schulten1999}, Eq. (5.226). Any such matrix has two opposite real eigenvalues $\mu ,-\mu $, and admits an orthogonal basis of eigenvectors, respectively $u \in {\sf C}^2$ for $\mu $ and $v$ for $-\mu $. Therefore, the $4\times 4$ matrix $P$ has the two same opposite real eigenvalues $\mu ,-\mu $ as has $N$. \\

\subsection{Completing the proof for the standard Dirac equation (DFW)}

Note that, in the evaluation of $\delta $E (Subsects. \ref{E explicit-DFW} and \ref{chiral}), we did not use the assumption that the local similarity is a small perturbation. Using this assumption, we have Eq. (\ref{delta lambda}). Using the latter, Eq.  (\ref{deltaE-final}), and the definition (\ref{Hermitian-sigma=1-g}) of the Hilbert space scalar product, which we rewrite as
\be \label{Hermitian-sigma=1-g-2}
(\psi  \mid \varphi  ) \equiv \int_{{\sf R}^3} \psi^\dagger B  \varphi \ \ \sqrt{-g}\ \dd^ 3{\bf x},
\ee
and in which $B$ is given by (\ref{a^0_j=0case}), we get the shift of the eigenvalue as
\bea\label{delta lambda DFW}
\delta \lambda & = & \int_{{\sf R}^3} \psi^\dagger P  \psi \ \sqrt{-g\,g^{00}}\ \dd^ 3{\bf x}\nonumber\\
& = & \int_{{\sf R}^3}(\phi ^\dagger N \phi + \chi ^\dagger N \chi)\sqrt{-g\,g^{00}}\ \dd^ 3{\bf x}\nonumber\\
& \equiv & \int(\phi ^\dagger N \phi + \chi ^\dagger N \chi)\ \dd {\sf V}', 
\eea
where we have decomposed the four-component eigenstate $\psi $ into two two-component states: $\psi =(\phi ,\chi )$. Assume (without any loss of generality) that, say, 
\be\label{phi>chi}
\int\phi ^\dagger \phi\ \dd {\sf V}' \geq  \int \chi ^\dagger \chi\ \dd {\sf V}'.
\ee
Select a constant $\mu >0$. Then, for each $x$ in the space manifold M ($x$ depends on the three coordinates $x^j,\ j=1,2,3$), we may select $N=N(x)$ in Eq. (\ref{deltaE-final}) such that $\phi (x)$ is an eigenvector of $N(x)$ with eigenvalue $\mu$, so that
\be
\phi ^\dagger N \phi=\mu\ \phi ^\dagger \phi.
\ee
Since $N$ has the eigenvalues $\mu $ and $-\mu $, we have then automatically
\be\label{chi vp -mu}
\chi ^\dagger N \chi \geq -\mu \ \chi ^\dagger \chi.
\ee
Here equality is got only if $\chi (x)$ is an eigenvector of $N(x)$ with eigenvalue $-\mu $.  In that case $\phi (x)$ is orthogonal to $\chi (x)$ in  ${\sf C}^2$.  Thus, we conclude that  $\delta \lambda >0$ in (\ref{delta lambda DFW}), unless $\phi (x)$ and $\chi (x)$ are orthogonal almost everywhere {\it and}  simultaneously $\int\phi ^\dagger \phi\ \dd {\sf V}' =  \int \chi ^\dagger \chi\ \dd {\sf V}'$. \\

Already the first feature is impossible if $m \ne 0$. It is well known that the probability current $J^\mu $ satisfies the following Fierz identity \cite{Pauli1936}:
\be
J^\mu J_\mu =\left(\psi ^\dagger A\gamma ^\mu \psi \right)\left(\psi ^\dagger A\gamma _\mu \psi \right)=\left(\psi ^\dagger A\psi  \right)^2+\left(i\psi ^\dagger A\gamma ^5 \psi \right)^2.
\ee
The right-hand side vanishes if $\phi (x)$ and   $\chi (x)$ are orthogonal in ${\sf C}^2$.  In that case, the probability current $J^\mu$ lies on the light cone (almost everywhere), which is impossible \cite{ReiflerMorris2005} for a solution of a Dirac equation with a nonvanishing mass term.  Q. E. D.   \\

\subsection{The case with TRD equations}\label{nonuniqueness-TRD}

Using (\ref{Q^croix B}), the difference operator (\ref{S Etilde S^-1 -E}) is rewritten, similarly as we did for DFW in Eq. (\ref{SEtilde-ES-DFW-2}), as
\bea\label{S Etilde S^-1 -E-2}
\delta \mathrm{E}'\equiv S\widetilde{\mathrm{E}}'S^{-1}-\mathrm{E}'& =& iB^{-1}\left[B^\mu(D_\mu S)S^{-1}-B(\partial _0S)S^{-1}\right]^a\nonumber\\
& = & i\left[B^{-1}B^\mu (D_\mu S)S^{-1}-(\partial _0S)S^{-1}\right]^{a(B)}.
\eea
Assuming again for simplicity that the matrix $B$ is proportional to the identity, as in Eq. (\ref{a^0_j=0case}) for DFW, this reduces to
\be\label{S Etilde S^-1 -E-3}
\delta \mathrm{E}'= i\left[B^{-1}B^\mu (D_\mu S)S^{-1}-(\partial_0 S)S^{-1}\right]^a
= i\left\{\left[(D_0 S)-(\partial _0S)+\alpha ^j(D_jS)\right]S^{-1}\right\}^a.
\ee
The r.h.s. of (\ref{S Etilde S^-1 -E-3}) shows that the operator $\delta \mathrm{E}'$ is multiplication by a $4\times 4$ Hermitian matrix which we may denote as $P$. For TRD, we do not have any restriction on the similarity $S$. We can set $\delta \mathrm{E}'\equiv P=iQ^a$ in Eq. (\ref{S Etilde S^-1 -E-3}). Then the matrix $Q$ can be chosen freely, by taking the similarity $S$ to solve the linear PDE
\be\label{S Etilde S^-1 -E-4}
D_0 S-\partial _0S+\alpha ^jD_jS=QS.
\ee
Hence, the Hermitian matrix $P=iQ^a$ can be chosen arbitrarily.  Then, the reasoning is simpler than for DFW: we get again $\delta \lambda >0$ by choosing, for each $x$, the matrix $P(x)$ in such a way that the eigenfunction $\psi $ of E takes a value $\psi (x)$ which is an eigenvector of  $P(x)$ for always the same eigenvalue $\mu >0$, for any $x \in \,$M. In fact, in that case, the unperturbed eigenfunction $\psi $ is an eigenfunction of the perturbed energy operator, with the eigenvalue $\lambda +\mu $.

\section{Conclusion}\label{Conclusion}

In the \hyperref[Introduction]{Introduction}, we asked three questions: {\bf i}) When does a change of the coefficient fields lead to an equivalent Hamiltonian operator? Or {\bf ii}) to an equivalent energy operator? {\bf iii}) Is the energy spectrum unique? [Note that, in experiments, observing different spectra is easier than determining whether two energy operators are equivalent, so that (iii) is stronger than (ii).]\\

\noindent {\bf i}) In Section \ref{EquivalentHamiltonians} we showed, for DFW, that the Hamiltonian operators H and $\widetilde {\mathrm{H}}$, before and after application of a local similarity $S$, are equivalent iff $S$ is independent of time.  For the two versions of TRD, the conditions are less simple, but only very special local similarities will satisfy them.  Thus, for both DFW and TRD, the Hamiltonian operator associated with the free Dirac equation in a curved spacetime is not unique. \\

\noindent {\bf ii}) Similar results were obtained in Section \ref{Equivalent-E-Operators} for the energy operator which, unlike the Hamiltonian operator, is always Hermitian with respect to the relevant Hilbert space scalar product as required for an observable. \\

\noindent {\bf iii}) For both DFW and TRD, we proved in Section \ref{Non-uniquenessProof} assuming very mild conditions that, for any admissible set of coefficient fields $(\gamma ^\mu ,A)$, there exists an admissible local similarity $S$ such that the spectra of the energy operators before and after the similarity are different.\\  

\vspace{3mm}
\noindent The non-uniqueness of the energy operator is significant because, according to quantum mechanics: if a particle is in a state $\psi $, then an energy measurement at any given time will yield an eigenvalue of the energy operator (at that time in the non-stationary case). Immediately following the measurement, the state $\psi $ will collapse into an eigenvector of the energy operator at that time, as a result of the measurement. Since the Hamiltonian and energy operators are not unique, and since there is no preferred way to choose them in either DFW or TRD, we cannot at present predict the energy of a free spin-half particle in a curved spacetime. However, the energy shifts due to different ``reasonable'' choices of the coefficient fields $(\gamma^\mu,A)$  are very small.  For example, for rotating Dirac matrices in a Minkowski spacetime, a straightforward calculation using Eqs. (\ref{deltaE}) and (\ref{e_ab s^ab ^antisym-2}) shows that the energy shift is of the order of the Mashhoon energy term \cite{Mashhoon1988}.  That means for the current upper limit of angular rotation rates of neutron stars, $10^3$ Hz, the energy shift is two orders of magnitude less than the experimental accuracy of the Lamb shift, which is $10^5$ Hz \cite{Ryder1996}. Nevertheless, in the current state of the theory, nothing prevents one from choosing a tetrad field with an arbitrarily large rotation rate, without any relation to the rotation of a physical body. 
\footnote{\
After the first version of the present work had been posted \cite{A43-v1}, an interesting attempt to solve the non-uniqueness problem of the Dirac Hamiltonian was proposed by Gorbatenko \& Neznamov \cite{GorbatenkoNeznamov2010,GorbatenkoNeznamov2011} for DFW, by restricting the class of tetrad fields. However, these authors do not address spectral inequivalence, which is a stronger condition than mere operator inequivalence.  Two energy operators may not be equivalent as operators, yet they may be isospectral, leading to the same physical observations of the energy. In the present paper, our unique aim has been to show in detail that there is a non-uniqueness problem for the Dirac Hamiltonian and energy operators and to demonstrate that it implies physically observable differences in the energy spectrum.}
\\

\end{document}